\documentclass[a4paper,11pt]{article}
\usepackage{jcappub}

\usepackage{amsmath}
\usepackage{amssymb}
\usepackage{amsfonts}
\usepackage{mathtools}

\usepackage{lineno}
\usepackage{lscape}
\usepackage{soul}
\usepackage{subfigure}
\usepackage{graphicx}
\usepackage{color}
\usepackage{bm}
\usepackage{epstopdf}
\usepackage{url}
\usepackage{booktabs}
\usepackage{mdframed}
\usepackage{hyperref}
\usepackage{booktabs}
\usepackage{siunitx}
\usepackage{multirow}
\usepackage{appendix}

\usepackage{changes}
\usepackage[colorinlistoftodos]{todonotes}

\usepackage[utf8x]{inputenc}
\usepackage[normalem]{ulem}

\definecolor{burgundy}{rgb}{0.6, 0.0, 0.0}
\definecolor{persianblue}{rgb}{0.11, 0.22, 0.73}
 \hypersetup{colorlinks, citecolor=burgundy, linkcolor=persianblue, urlcolor=persianblue}
 \hypersetup{colorlinks, citecolor=burgundy, linkcolor=persianblue, urlcolor=persianblue}

\definecolor{Gray}{gray}{0.95}

\begin{document}
\title{Multi-wavelength astronomical searches for primordial black holes}

%\title{Multi-wavelength astronomical searches for primordial black holes: A comprehensive analysis.}

\author[a,b]{Julien Manshanden}
\emailAdd{julien.manshanden@desy.de}
\author[a,c]{Daniele Gaggero}
\emailAdd{daniele.gaggero@uam.es}
\author[a]{Gianfranco Bertone}
\emailAdd{g.bertone@uva.nl}
\author[d]{Riley M. T. Connors}
\emailAdd{rconnors@caltech.edu}
\author[e]{Massimo Ricotti}
\emailAdd{ricotti@astro.umd.edu}

\affiliation[a]{\small GRAPPA, Institute of Physics, University of Amsterdam, 1098 XH Amsterdam, The 
Netherlands}
\affiliation[b]{\small II. Institut für Theoretische Physik, Universität Hamburg, 22761 Hamburg, Germany} %it's the second theoretical institute
\affiliation[c]{\small Instituto de F\'isica Te\'orica UAM-CSIC,
Campus de Cantoblanco, E-28049 Madrid, Spain}
\affiliation[d]{\small Cahill Center for Astronomy and Astrophysics, California Institute of Technology, Pasadena, CA 91125, USA}
\affiliation[e]{\small Department of Astronomy, University of Maryland,
College Park, MD 20740, USA}

\date{\today}

%\begin{abstract}
%{
%Primordial black holes of $\mathcal{O}(1-100) M_{\odot}$ can constitute a significant portion of the dark matter in the Universe. Given the large amount of interstellar gas in our Galaxy, these objects can accrete gas and shine in the radio and X-ray bands. In this paper we present a comprehensive discussion of the upper limit on the PBH abundance based on the comparison with known radio and X-ray sources in the Galactic center region. We consider the impact of extended mass functions on the bound, showing that it becomes stronger with respect to the case of a monochromatic mass function. In the second part we place the bound on more solid physical grounds by taking into account the results of a set of state-of-the-art numerical simulations designed to capture the physics of gas accretion onto isolated, moving compact objects. We show that the bound gets two orders of magnitude stronger with respect to previous results based on a conservative phenomenological treatment of the accretion physics: Therefore, the current upper limit to the PBH abundance in the mass range of interest is set to $\sim 10^{-3}$ of the dark matter density. We provide in the last part of the paper a comprehensive critical discussion on the reliability of this bound and possible future developments in the field.
%}
%\end{abstract}

\begin{abstract}
{
If primordial black holes of $\mathcal{O}(1-100) M_{\odot}$ constitute a significant portion of the dark matter in the Universe, they should be very abundant in our Galaxy. We present here a detailed analysis of the radio and X-ray emission that these objects are expected to produce due to the accretion of gas from the interstellar medium. With respect to previous studies, we relax the assumption of a monochromatic mass function, and introduce an improved treatment of the physics of gas accretion onto isolated, moving compact objects, based on a set of state-of-the-art numerical simulations. By comparing our predictions with known radio and X-ray sources in the Galactic center region, we show that the maximum relic density of primordial black holes in the mass range of interest is $\sim 10^{-3}$ smaller than that of dark matter. The new upper bound is two orders of magnitude stronger with respect to previous results, based on a conservative phenomenological treatment of the accretion physics. We also provide a comprehensive critical discussion on the reliability of this bound, and on possible future developments in the field. We argue in particular that future multi-wavelength searches will soon start to probe the galactic population of astrophysical black holes.
}
\end{abstract}

%The new upper bound is two orders of magnitude stronger with respect to previous results based on a conservative phenomenological treatment of the accretion physics. It is also close to the expected density of astrophysical black holes in our Galaxy, and the search for that population is a major challenge in itself. Finally we provide a comprehensive critical discussion on the reliability of this bound, and on possible future developments in the field.

\keywords{primordial black holes, dark matter, radio astronomy, gravitational wave astronomy}

\maketitle

%\linenumbers

\section{Introduction}

{The idea that primordial black holes (PBHs) exist in the Universe \cite{1967SvA....10..602Z,Hawking:1971ei,Carr:1974nx} has been widely discussed in the past few decades~\cite{Carr:2016drx}}. These hypothetical objects could have formed before big-bang nucleosynthesis out of the collapse of small-scale large-amplitude density fluctuations originated during inflation, or through a variety of other mechanisms including phase transitions, topological defects such as cosmic strings and domain walls, condensate fragmentation, and bubble nucleation (see e.g. Refs. \cite{Green:2014faa,Sasaki:2018dmp}). 

PBHs with mass between $\sim 10^{-18}$ and $\sim 10^6$ solar masses have been invoked as possible dark matter candidates  \cite{Hawking:1971ei,ChaplineNature1975}. One of the most interesting mass window appears to be the one associated with the massive black-hole-binary merger events reported by the LIGO and VIRGO collaborations, centered on $\mathcal{O}(10)$ solar masses. These events  have indeed significantly revived the interest in this topic~\cite{Bird:2016dcv,Clesse:2016vqa,Kashlinsky:2016sdv}, encouraged a re-evaluation of the existing bounds~\cite{Ricotti:2007au,Carr:2016drx,Garcia-Bellido:2017xvr}, and triggered an extended discussion on the prospects for astronomical and cosmological signatures~\cite{Garcia-Bellido:2017fdg,Clesse:2017bsw}. 

% we focus our attention on the prospects for detecting a population of PBHs in our Galaxy by means of multi-wavelength astronomical observations. 

%There are several motivations for studies of this kind, in a multi-messenger context: On the one hand, the future appears very promising as far as the detection of gravitational waves is concerned, given the existing and planned facilities in different regions of the world; on the other hand, the radio astronomy field will benefit from an unprecedented increase in the sensitivity thanks to the SKA project, an array of detectors that will provide a sharp view of the radio sky from from $\simeq 50$ MHz to $\simeq 14$ GHz over the next decade.

In a previous paper \cite{Gaggero:2016dpq} we have shown that the hypothesis that PBHs comprise all the DM in the universe can be constrained by exploiting Galactic astronomical data, in the radio and X-ray bands: The idea is that a small fraction of these objects are expected to accrete interstellar gas in a significant way, especially in the inner Galactic bulge, and therefore show up as radio and X-ray sources in the sky. By  comparison with current source catalogs we could place a conservative bound on their abundance in the $10$ - $100$ $M_{\odot}$ mass window. 

%However, this bound still allows a non-negligible fraction of the DM to be in the form of PBHs. 

In view of the upcoming observing runs of advanced gravitational waves interferometers, and in preparation for the exquisite radio observations that will become available thanks to the SKA project -- an array of detectors that will provide a sharp view of the radio sky from from $\simeq 50$ MHz to $\simeq 14$ GHz -- it is important to go beyond the conservative approach of Ref.~\cite{Gaggero:2016dpq}, and to address in detail a number of important issues like the dependence of the bound on the mass distribution function, and the uncertainties inherent to the accretion physics. 

The goal of this paper is precisely to go beyond the conservative estimates presented in Ref.~\cite{Gaggero:2016dpq}. We will first present a robust and updated upper bound on the DM fraction in PBHs that takes into account the possibility that PBHs have a general `extended' mass function. We will then implement state-of-the-art numerical simulations of gas accretion onto moving BHs in order to more carefully characterize the expected multi-wavelength emission of PBHs, and to put the upper bound on the contribution of PBHs to the dark matter abundance itself on solid physical grounds.

%\section{Ideas/structure for the Paper}

%Results we want to include:
%[We improve upon the results of Gaggero et al. 2017] [1) revising accretion formalism (using Park\&Ricotti 2013 without bursts to be more conservative)] [2) implementing velocity dist (need to include baryonic contribution?)]
%[EMD in appendix? (to show the effect on the new constraints)]

%Thus plots for 1 only, 2 only and 1 and 2 combined?

%To do 2 I need a way to fix the velocity distribution for the baryonic contribution. Using 

\section{Impact of the mass distribution on the astronomical bounds.}
\label{sec:impact_mass}

\subsection{Introductory considerations}
\label{subsec:intro_con}

The expected mass distribution of PBHs has been a subject of much discussion since the first pioneering papers from Carr and Hawking
(see, for instance, \cite{1975ApJ...201....1C}). 
%As noticed many times in the more recent literature, a very narrow mass distribution, well approximated by a Dirac delta, is actually very unlikely in most inflationary models that may produce a significant amount of PBHs.  
Although narrow mass distributions, well approximated by a Dirac delta, could be produced e.g. in the context of single-field inflation models \cite{Garcia-Bellido:2017mdw,Ezquiaga:2017fvi,Ballesteros:2017fsr,Cicoli:2018asa,Ozsoy:2018flq},
extended mass distributions (EMD) can arise from a variety of mechanisms: For instance, broad log-normal mass distributions peaked at large stellar masses can arise within hybrid inflation scenarios \cite{Clesse:2015wea}; power-law distributions, on the other hand, can arise within inflation models featuring either a spectator field with a blue spectrum or a running spectral index \cite{Carr:2017edp}. A variety of other mechanisms, such as phase transitions \cite{PhysRevD.26.2681,Rubin:2001yw}, or collapse of cosmic strings \cite{Hawking1989237} or other topological defects, can also generate black holes with extended mass functions. 
 
The impact of EMDs on constraints on the PBH abundance was first analyzed in \cite{Carr:2016drx,Green:2016xgy,Carr:2017jsz}. An improved remapping method to convert the constraints was later proposed in \cite{Bellomo:2017zsr} and applied to lensing and CMB constraints.

Given these considerations, it is very useful to analyze the impact of different types of EMDs on the radio and X-ray bounds reported in \cite{Gaggero:2016dpq}. 
Following the approach of that paper, we consider a set of Monte Carlo simulations featuring a population of PBHs that trace the DM distribution in the Galaxy. The simulations include a realistic distribution of the interstellar gas, and rely on conservative assumptions regarding the physics of accretion and the subsequent non-thermal emission in the radio and X-ray bands: We refer to Ref.~\cite{Gaggero:2016dpq} for the details about the procedure and the astrophysical ingredients, here we recall the main features:

\begin{itemize}

\item[$-$] We assume that the PBHs follow a Navarro-Frenk-White profile \cite{Navarro:1995iw} (for a discussion of the impact of this assumption, see Sec. 4), and the gas distribution in the inner Galaxy is modeled as in \cite{Ferriere2007}. 

{
\item[$-$] We assume that the PBH velocity
 distribution at a distance $R$ from the Galactic Center is a Maxwell-Boltzmann with mean velocity
 \begin{equation}
      v_{\rm mean} = v_{\rm circ}(R) = \sqrt{\left( G \, M(<R) / R \right)}.
 \end{equation}
 For the mass distribution in the Milky Way, $M(R)$, we consider the recent state-of-the-art model taken from~\cite{McMillan2016}.
}

\item[$-$] {We conservatively assume, following, e.g., Ref. \cite{Fender:2013ei}, (see Section 3 for a more complete discussion of the accretion physics) that the accretion rate of isolated PBHs is a small fraction $\lambda$ of the Bondi-Hoyle-Littleton rate,
\begin{equation}
    \dot{M} = \lambda \dot{M}_\mathrm{BHL}, \label{eq:old_accretion}
\end{equation}
with,}
\begin{equation}
    \dot{M}_\mathrm{BHL} \equiv 4 \pi (G M_\mathrm{BH})^2 \rho (v^2 + c_\mathrm{s}^2)^{-3/2} \label{eq:BHL_accretion}
\end{equation}
where $G$ is the gravitational constant, $M_\mathrm{BH}$ is the BH mass, $\rho$ is the density of the ambient gas, $v_\mathrm{BH}$ is the velocity of the BH (expressed here as a simple scalar quantity), and $c_\mathrm{s}$ is the sound speed of the gas. The adopted value $\lambda = 0.02$ is consistent (albeit approaching the upper bounds) with isolated neutron star population estimates and studies of active galactic nuclei accretion~\cite{Perna:2003,Pellegrini:2005,wang2013}.

%\begin{eqnarray*}
%    \dot{M}_\mathrm{BHL} &\equiv& \frac{\rho_\mathrm{in} %(GM_\mathrm{BH})^2}{(v_\mathrm{in}^2 + c_\mathrm{s,in}^2)^{3/2}} \\
%    \dot{M} &=& \lambda \dot{M}_\mathrm{BHL}
%\end{eqnarray*}

%\begin{equation}
%\dot{M}=4\pi\lambda (G M_\mathrm{BH})^2 \rho %\left(v_\mathrm{BH}^2+c_\mathrm{s}^2\right)^{-3/2}\,\,\, ,
%\label{eq:accretion}
%\end{equation}

\item[$-$] We model the radiative efficiency $\eta$ as follows:

\begin{equation}
L_B = \eta \dot{M} c^2
\label{eq:bolometricL}
\end{equation}

with $\eta = 0.1 \dot{M}/\dot{M}{\rm_{crit}}$ for $\dot{M}<\dot{M}\rm_{crit}$. {The critical accretion rate $\dot{M}_\mathrm{crit}$ is defined in terms of the Eddington luminosity $L_\mathrm{Edd}$ as 

%$\dot{M}_{\rm crit} \equiv 0.01~\dot{M}\rm_{Edd}$.

%{\bf The critical accretion rate is defined as %$\dot{M}_{\rm crit} \equiv 0.01~\dot{M}\rm_{Edd}$.

%\begin{equation}
%\dot{M}_{\rm crit} \equiv \frac{L_{\rm Edd}}{\eta c^2}
%\end{equation}

%where $L_{\rm Edd}$ is the Eddington luminosity.}

\begin{equation}
\dot{M}_{\rm crit} \equiv 0.01 \cdot \frac{L_{\rm Edd}}{\eta_0 c^2},
\end{equation}

for which we fix the radiative efficiency to $\eta_0 = 0.1$, following, e.g., \cite{Fender:2013ei}.
} Since all the PBHs in our simulations fall below the critical accretion rate, they are inefficient accretors: As a consequence the luminosity scales non-linearly with the accretion rate: $L\propto\dot{M}^2$ (see, e.g., \cite{Narayan:1994,heinz2003} for more details about this regime). 

\item[$-$] We compute the number of black holes in our simulations that would appear as point sources in the radio (in the GHz domain) and in the soft X-ray band (0.5-8 keV)  above the detection threshold of the VLA and Chandra experiments respectively. To do so, we assume that $30\%$ the bolometric luminosity estimated by means of Eq. \ref{eq:bolometricL} is radiated in the soft X-ray band, and we convert the X-ray to radio luminosity by applying the empirical {\it fundamental plane} (FP) relation \cite{2012MNRAS.419..267P}, which applies for a large class of
compact objects in different mass ranges, from X-ray binaries all the way up to active Galactic nuclei. We then compare the predicted number of radio and X-ray sources associated with PBHs in our simulations to the number of observed ones, in a small region of interest centered on the Galactic center (see \cite{Gaggero:2016dpq} for the full detail) dominated by a massive molecular structure known as the {\it Central Molecular Zone} (CMZ), and place the bound on the abundance of PBHs accordingly.

\end{itemize}

In the following section we discuss how the results in Ref.~\cite{Gaggero:2016dpq} change when adopting an extended PBHs mass distribution $f(M)$. 

\subsection{Log-normal mass function}

We start by considering a log-normal distribution written as:

\begin{equation}
%\bbox{
f(M | \mu, \sigma) \,=\, \frac{1}{M \sqrt(2 \pi) \sigma} \exp{\left(  - \frac{log(M/\mu)^2}{2\sigma^2} \right)},
%}
\end{equation}
where $M$ is the random variable, $\mu$ is the {\it median} mass, and $\sigma$ is the standard deviation of the distribution in the $log(M)$ space. The normalization is set to 1 so that the actual mass distribution is given by $dN/dM = N_{\rm tot} f(M)$.

\begin{figure}[h]
\centering
\includegraphics[width=7.5cm]{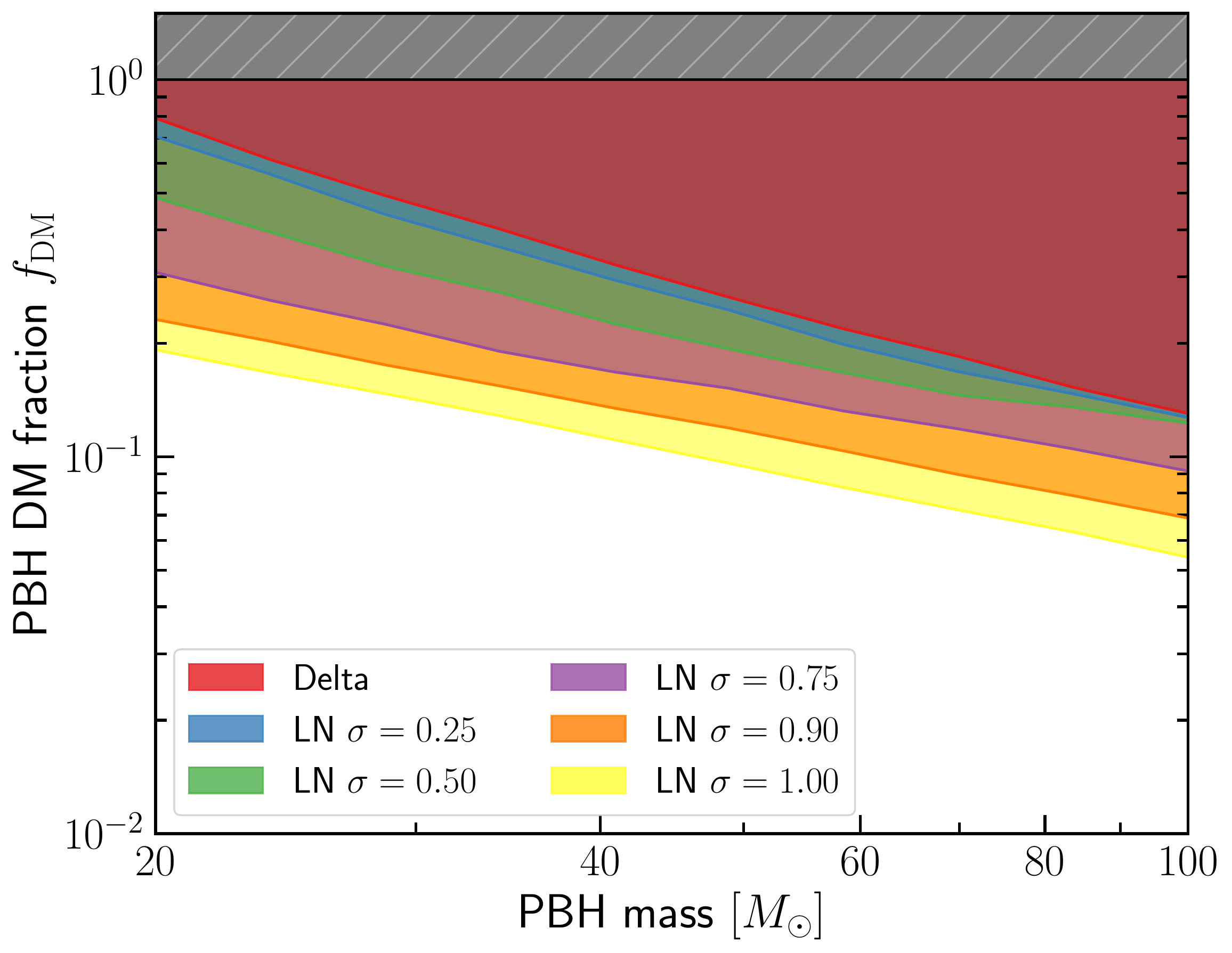}
\includegraphics[width=7.5cm]{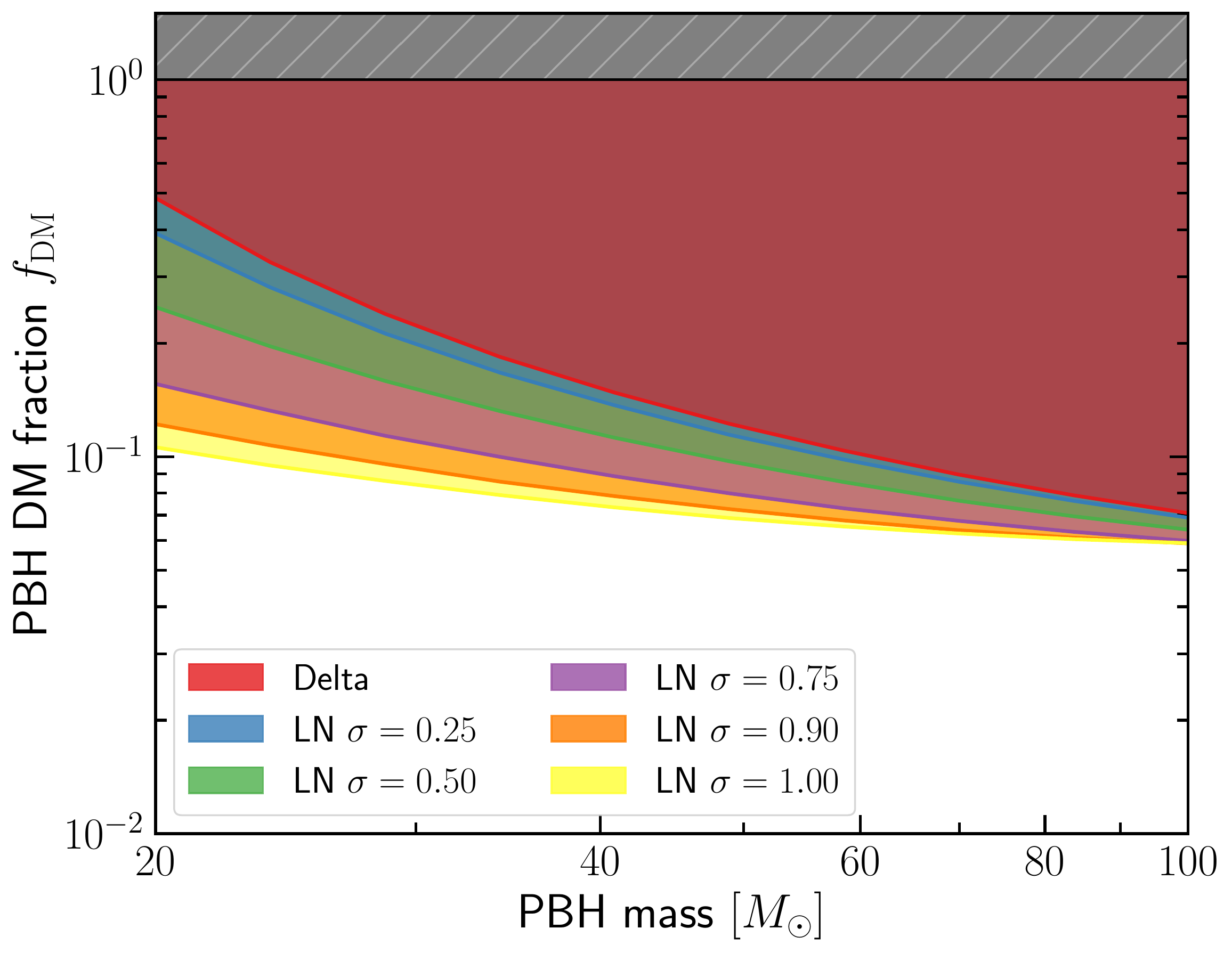}
\caption{$5\sigma$ constraints on the PBH DM fraction for different mass distributions, obtained by comparison with radio (left) and X-ray (right) observations. The delta mass functions (red) have a parameter equal to the PBH mass given on the x-axis. The remaining mass functions are log-normal distributions with median given by the PBH mass on the x-axis. The sigma parameters of the log-normal distributions, indicating the spread of the distribution, are taken as $\sigma=0.25$ (blue), $\sigma=0.50$ (green), $\sigma=0.75$ (purple), $\sigma=0.90$ (orange), and $\sigma=1.00$ (yellow).}
\label{fig:lognormal}
\end{figure}

The results associated with this distribution are visualized in Fig. \ref{fig:lognormal}. 
In principle, since the constraints depend on the {\it number of detectable black holes} (i.e., the black holes emitting radio or X-ray radiation above the detection threshold), and given that PBHs with mass $M > \mu$ ($M < \mu$) get higher (lower) fluxes with the implementation of an extended instead of a monochromatic (i.e., Dirac delta) mass distribution, the bound could get stronger or weaker depending on how the competing effects on the high- and low-mass tails of the distribution balance.
In our case, given the fact that the constraints are driven by a tiny fraction of PBHs that---due to the low BH velocity and high ambient density---radiate above threshold, the implementation of log-normal mass distributions increases the fluxes of the PBHs close to the detection threshold, which in turn increases the number of detectable PBHs resulting in stronger constraints with increasing $\sigma$.

\subsection{Power-law mass function}

\begin{figure}[b!]
\centering
\includegraphics[height=6.5cm]{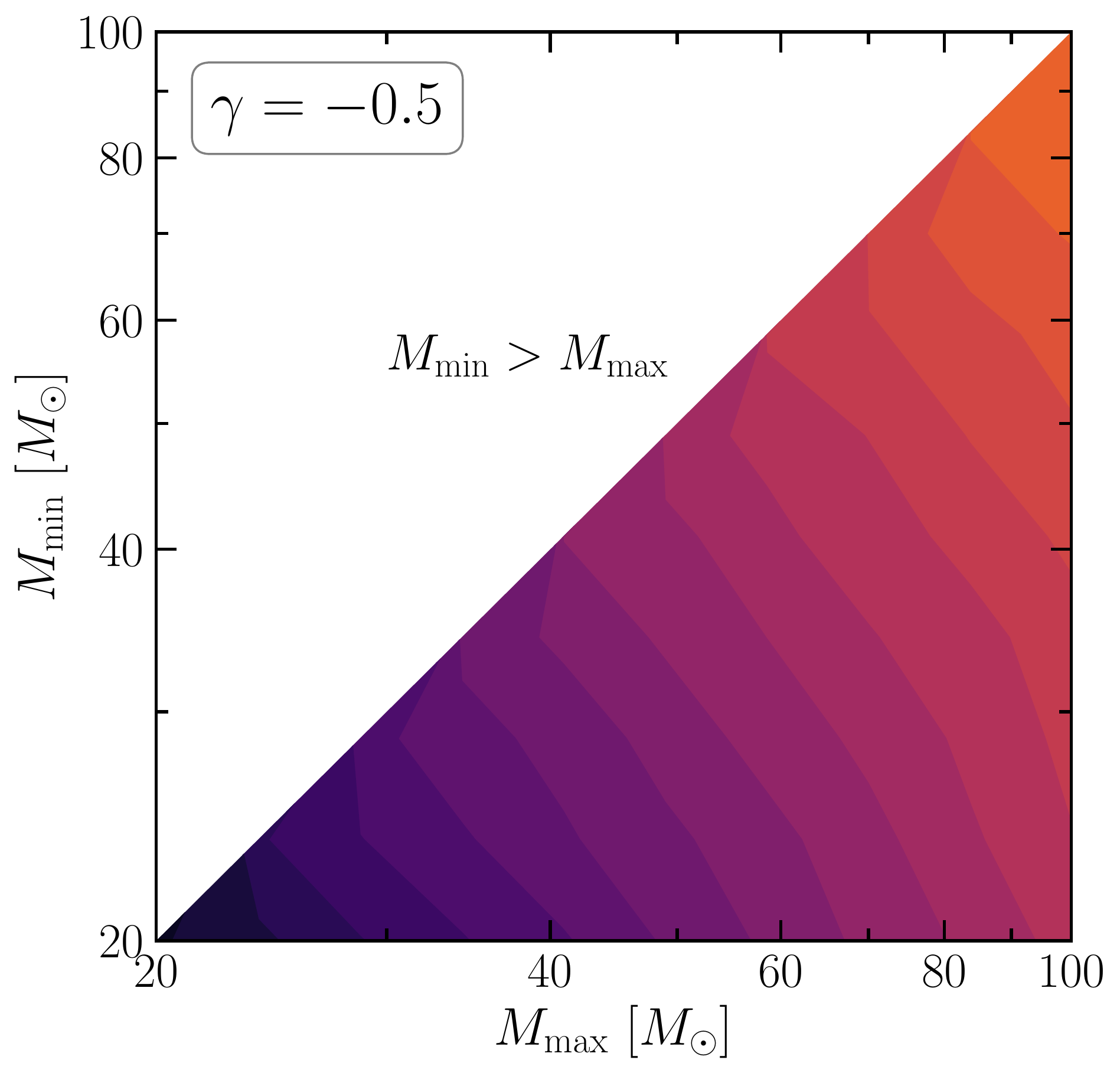}
\includegraphics[height=6.5cm]{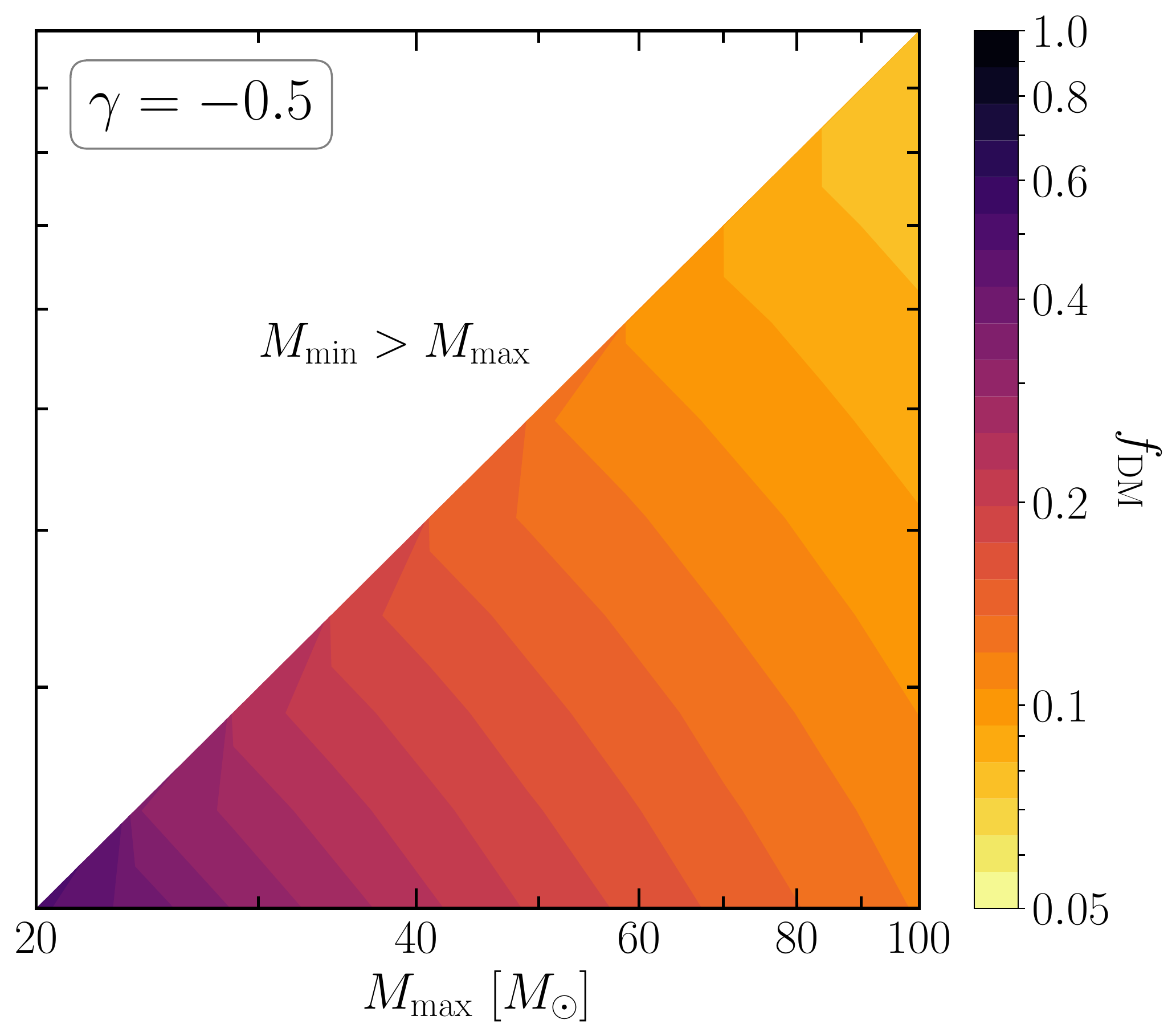}
\includegraphics[height=6.5cm]{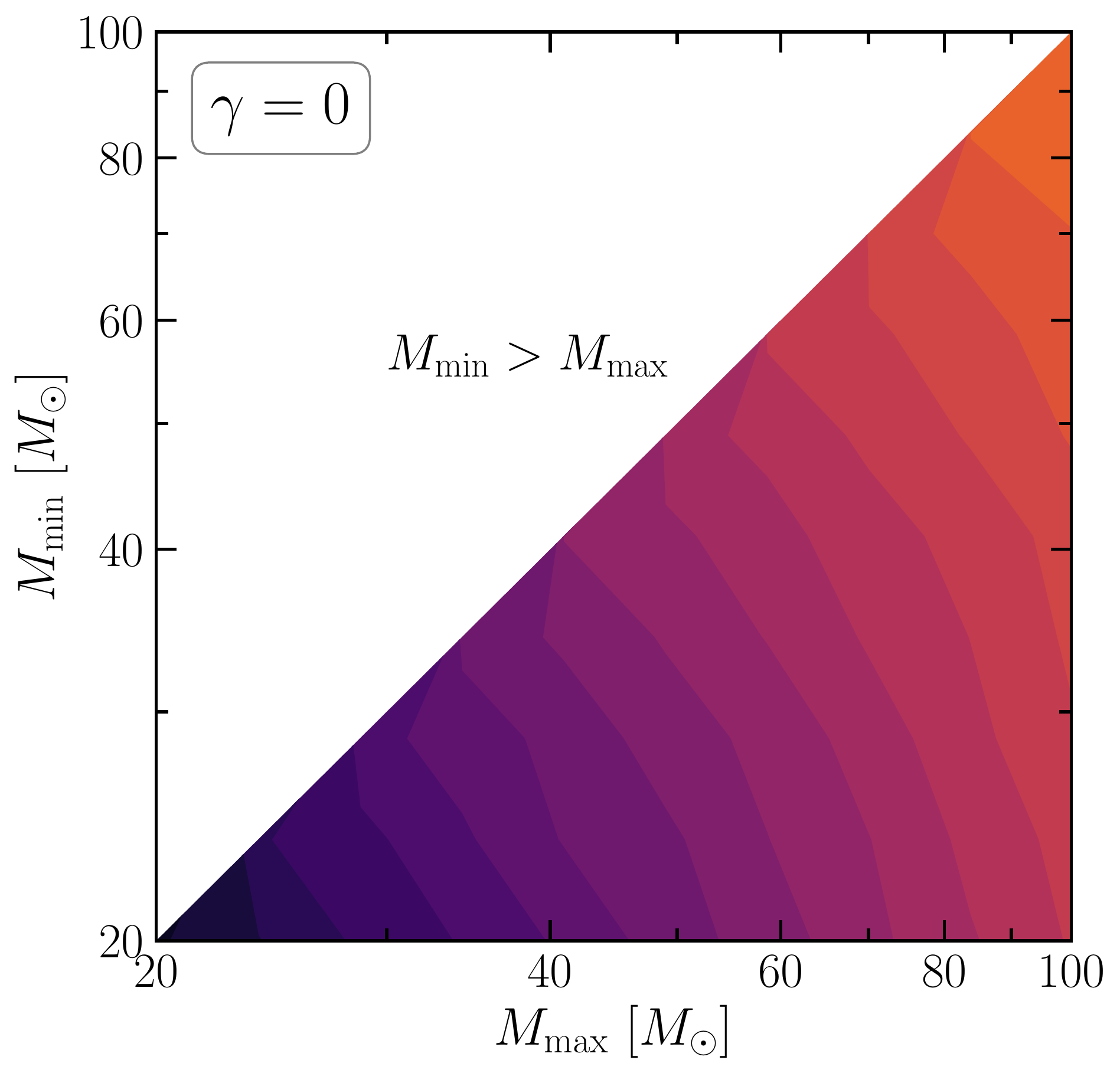}
\includegraphics[height=6.5cm]{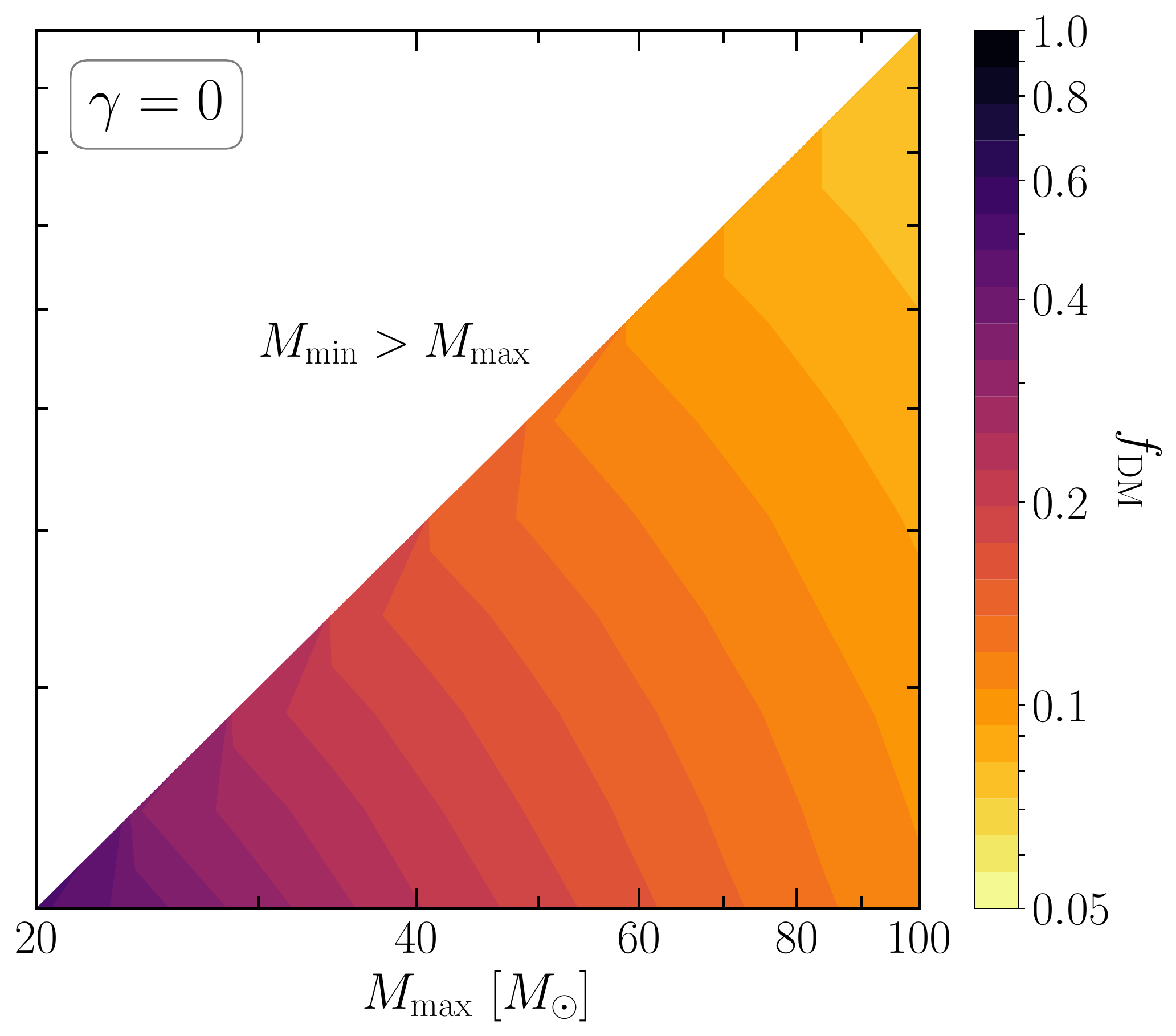}
\caption{$5\sigma$ constraints on the PBH DM fraction for different parameters of the power-law mass distribution. For each mass distribution, represented by a point in the ($M_\mathrm{max}$,$M_\mathrm{min}$)-space, the color represents the DM fraction $f_\mathrm{DM}$ that is excluded with a $5\sigma$ significance, as shown by the bar on the right. The different panels are for $\gamma = -0.5$ (top) and 0 (bottom), and radio (left) and X-ray (right) constraints. The region $M_\mathrm{min} > M_\mathrm{max}$ is not allowed and therefore left blank. }
\label{fig:powerlaw3D}
\end{figure}

Let us now turn our attention to power-law EMDs. We parametrize the power-law EMD (normalized to 1) as in \cite{Bellomo:2017zsr}:

\begin{equation}
%\bbox{
f(M | \gamma, M_{\rm min}, M_{\rm max}) \,=\, \frac{\mathcal{N}}{M^{1 - \gamma}} \Theta(M - M_{\rm min}) \Theta(M_{\rm max} - M),
%}
\end{equation}
where the Heaviside step functions $\Theta$ implement the low- and high-mass cutoffs, and the normalization factor $\mathcal{N}$ is given by

\begin{equation}
    \mathcal{N} \,=\,
    \begin{cases}
      \frac{1}{log(M_{\rm max}/M_{\rm min})}, & \text{if}\ \gamma = 0, \\
      \frac{\gamma}{M_{\rm max}^\gamma  - M_{\rm min}^\gamma}, & \text{if}\ \gamma \neq 0. \\
    \end{cases}
\end{equation}

Such distributions have been considered since the early works by Carr and Hawking (see e.g. \cite{Carr:1975qj} and \cite{Hawking1989237}), and is expected, e.g., if PBHs form out of the collapse of large density perturbations, or cosmic strings. We recall that the exponent $\gamma$ depends on the epoch of collapse. If the equation of state of the Universe at PBH formation is paramerized in the usual way as $P = w \rho$, the index $\gamma$ can be written as:

\begin{equation}
\gamma = -\frac{2 w}{1 + w}\,\, .
\end{equation}

For $w > -1/3$, as expected in an accelerating Universe, $\gamma$ covers the [-1, 1] range. Following  \cite{Bellomo:2017zsr}, we have considered two reference values: $\gamma = -0.5$, which corresponds to PBH formation during the radiation-dominated era ($w = 1/3$), and $\gamma = 0$, which corresponds to PBH collapse during matter domination ($w = 0$).
The X-ray and radio bounds based on the power-law distribution for these relevant values of the power-law index $\gamma$ are visualized in Fig.~\ref{fig:powerlaw3D}.
 The color code in that plot corresponds to the PBH abundance excluded at $5\sigma$, for both the radio (left), and X-ray (right) case: the excluded value can be as low as $5 \cdot 10^{-2}$ for large values of $M_{\rm max}$. We point out the very mild dependence on $M_{\rm min}$, and {\it vice versa} a strong dependence on $M_{\rm max}$. This behaviour is due to the fact that most of the constraining power actually comes from the high-mass tail, given the non-linear increase of the radiative output with $M$ and the subsequent significant increase of the number of high-mass PBHs emitting above threshold when $M_{\rm max}$ is increased.
 %\todo{I think this conclusion is a bit too fast: it depends strongly on Mmax because the high-mass tail gives rise to the constraints. That this is the case is indeed because the radiative output increases with mass, but the necessity of the non-linearity is not that trivial? I would be a bit more cautious here. I think the main reason why it is mildly dependent is that by increasing Mmin only a few heavier mass sources appear whereas an increase of Mmax causes a lot more heavy mass sources. {\color{red} DG:Changed} }

\subsection{A specific realization: PBH formation by vacuum bubbles.}

%\begin{figure}[h!]
%\centering
%\includegraphics[height=6.55cm]{powerlaw_radio_paper.pdf}
%\includegraphics[height=7cm]{powerlaw_xray_paper.pdf}
%\caption{$5\sigma$ constraints on the PBH DM fraction from the broken power law mass distribution. For each mass distribution, represented by a point in the ($M_*$,$M_\mathrm{min}$)-space, the color represents the DM fraction $f_\mathrm{DM}$ that is excluded with a $5\sigma$ significance, as shown by the bar on the right. Note that this color bar has a significantly different scale than in figure \ref{fig:powerlaw3D}. The different panels are for the radio (left) and X-ray (right) constraints. The region $M_\mathrm{min} > M_*$ is not allowed and therefore left blank. }
%\label{fig:paperpowerlaw3D}
%\end{figure}

Let us now consider in more detail, as a case study, a specific inflationary model that predicts a broad power-law spectrum of PBHs. Ref. \cite{Garriga:2015fdk} first noticed that non-perturbative quantum effects may play a relevant role during inflation. In particular, that paper proposed scenarios in which vacuum bubbles form during the inflationary phase due to the presence, in the underlying particle physics model, of {\it vacua} characterized by a lower energy density with respect to the one associated to the false vacuum that actually drives the inflationary expansion of the Universe. These bubbles are pulled inwards by the
negative pressure of vacuum in its interior, and may collapse to PBHs after inflation. In ref \cite{Deng:2017uwc}, a broken power-law mass spectrum is derived by means of numerical simulations. This spectrum extends over many decades in mass, and---given the many existing constraints---the total density of PBHs in this scenario is bound to be below $10\%$ of the dark matter density; interestingly, the number of predicted PBHs in the range covered by Virgo and LIGO are compatible with the rate of binary-black-hole mergers reported by those collaborations.

Inspired by these results, we consider a mass distribution of the form:

\begin{equation}
%\bbox{
    f(M) \,\propto\,
    \begin{cases}
      M^{-1}, & \text{if}\ M_{\rm min} < M < M_{\ast} \\
      M^{-3/2}, & \text{if}\ M > M_{\ast} \\
    \end{cases}
%    }
\end{equation}
We also introduce a cutoff at $M_{\rm max} = 1000$ $M_{\odot}$, given the very strong constraints on the PBH abundance in the high-mass domain (mainly from CMB measurements).% The results for the X-ray and radio bound are shown in Fig. \ref{fig:paperpowerlaw3D}. 
The $5\sigma$ constraint is strong and very weakly dependent on the parameters involved: The maximum allowed DM fraction in the form of PBHs in such a scenario lies within the range $[5 - 6] \cdot 10^{-2}$.\footnote{For ease of comparison with \cite{Deng:2017uwc}, this range translates to $[4-5] \cdot 10^{-2}$ and $[7-9] \cdot 10^{-3}$ at $2\sigma$ significance for X-ray and radio, respectively, thus providing an improvement to the existing constraints by a factor of 2 to 15.} %for 3$\sigma$ the ranges are [4.8-5.5] \cdot 10^{-2} and [1.7-2.1] \cdot 10^{-2} for X-ray and radio  %for 2$\sigma$ the ranges are [4.6-5.2] \cdot 10^{-2} and [7.8-9.4] \cdot 10^{-3} for X-ray and radio

\subsection{Remapping from monochromatic constraints.}
\label{subsec:remap}

It is now useful to consider the remapping method proposed in \cite{Bellomo:2017zsr} and compare that prescription to our results, as a further validation of our computations. 
The remapping procedure is based on the principle that different PBHs contribute independently to the observable of interest. Therefore, the observable of interest is, in general, proportional to a weighted integral over the mass distribution $d\Phi / dM$,\footnote{To follow the notation of \cite{Bellomo:2017zsr} and avoid confusion with the PBH dark matter fraction $f_\mathrm{PBH}$, here the mass distribution is denoted by $d\Phi/dM$ instead of $f(M)$. The meaning is exactly the same as before.}
\begin{equation}
f_{\rm PBH} \int { g(M, \{p_j\}) \frac{d\Phi}{dM} \,{ d} M },
\label{eq:observable}
\end{equation}
with the function $g(M,\{p_j\})$ encoding the underlying physics that determines the contribution of PBHs of mass $M$ to the observable. In our case, the observable of interest is the number of accreting black holes radiating above the detection threshold, with $g$ containing all the details regarding the physics of accretion.

Since Eq. \ref{eq:observable} holds for any mass function, one can compare to the case of a Dirac delta and write:
\begin{equation}
f_{\rm PBH}^{\rm MMD} \, g(M_{eq}, \{p_j\}) \,=\, f_{\rm PBH}^{\rm EMD} \int{ g(M, \{p_j\}) \frac{{ d} \Phi}{{ d} M} \,{ d}M  },
\end{equation}
where the superscript ${\rm MMD}$ refers to the monochromatic mass distribution. This equation implies that there exists a MMD, peaked at a given {\it equivalent mass} $M_{eq}$, that gives rise to the same value of the observable as the EMD under consideration: Therefore, the value of the constraint for an EMD can be read off from the delta-function constraint computed at $M_{eq}$. 

\begin{figure}[h!]
\centering
\includegraphics[width=7.5cm]{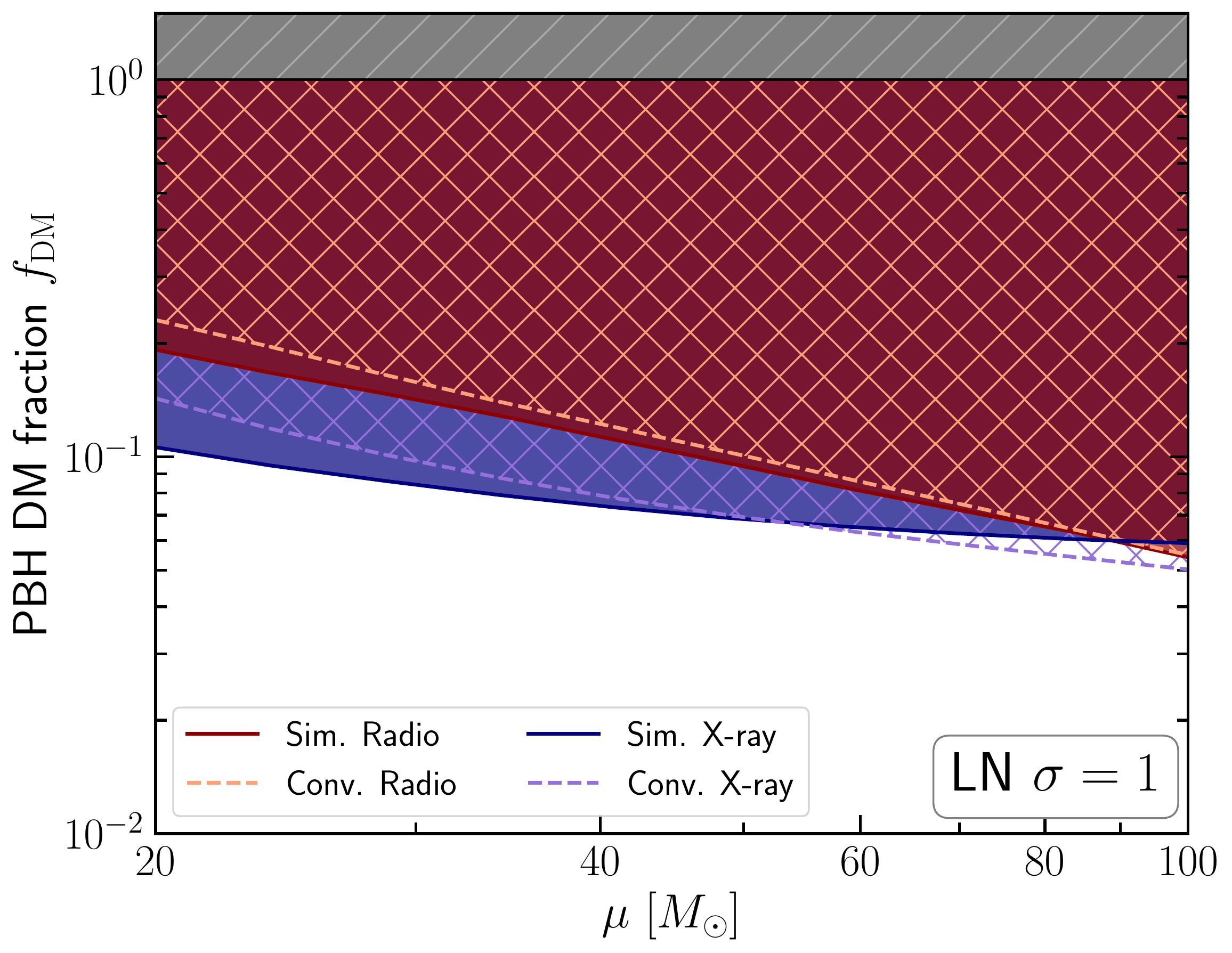}
\includegraphics[width=7.5cm]{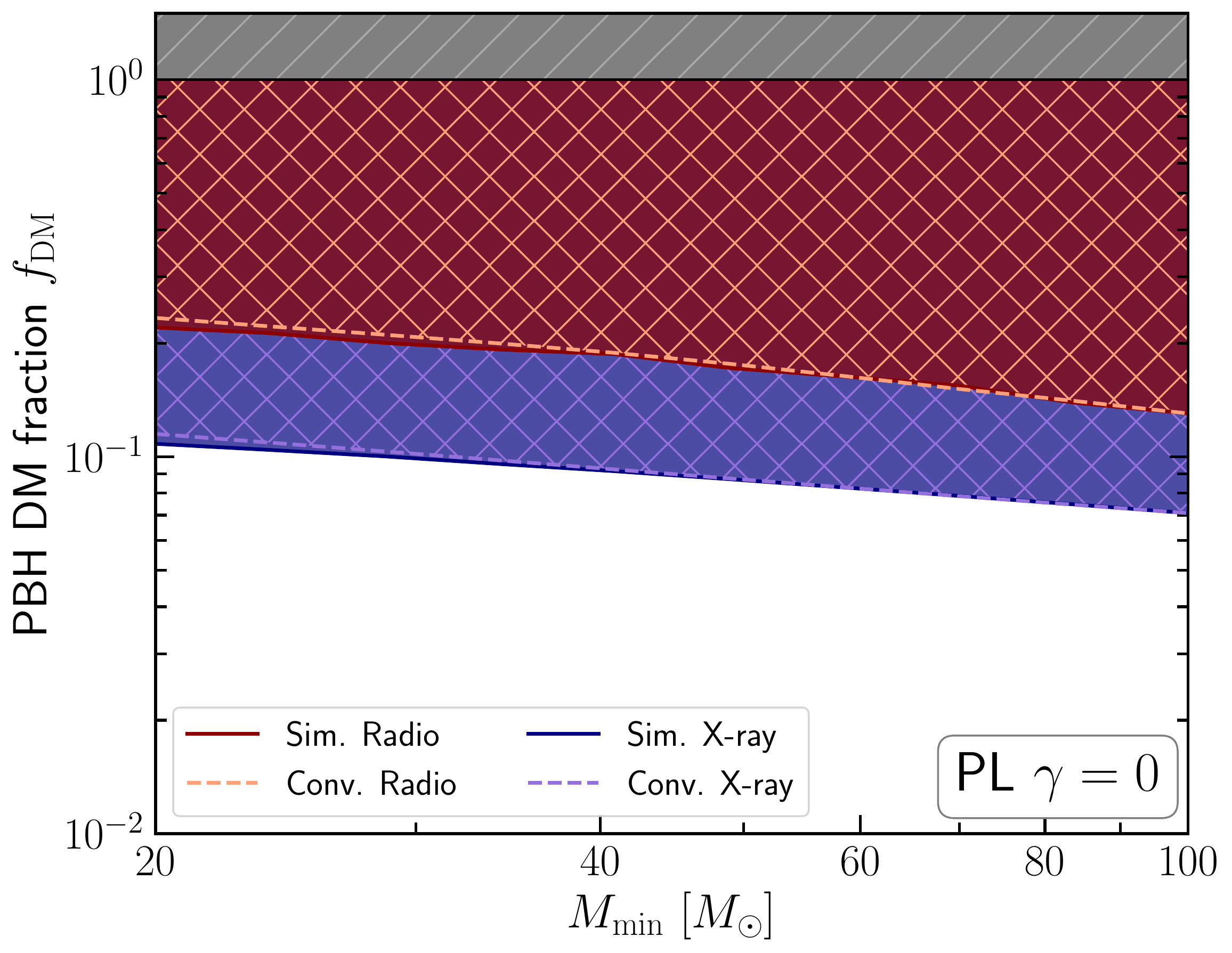}
\caption{Comparison of $5\sigma$ EMD constraints obtained from the implementation of EMDs in the simulations (solid, radio: dark red, X-ray: blue) with $5\sigma$ EMD constraints obtained from the conversion of MMD constraints (dashed and checkered, radio: salmon, X-ray: purple). The EMDs are log-normal distributions with $\sigma=1$ and $\mu$ varied on the x-axis (left) and power law distributions with $\gamma=0$, $M_\mathrm{max}=100M_\odot$ and $M_\mathrm{min}$ varied on the x-axis (right).}\label{fig:compareconv}
\end{figure}

We calculated the function $g(M, \{p_j\})$ for the X-ray and radio bound (see Appendix \ref{app:gfunction} for the details), and compared the converted constraints to our results obtained by means of Monte Carlo simulations. The comparison is shown in figure \ref{fig:compareconv}, the agreement is very good, and provides a solid, independent validation of our results.

%%%%%%%%%%%%%%%%%%%%%%%%%%%%%%%%%%%%%%%%%%%%%%%%%%%%%%%%%
%%%%%%%%%%%%%%%%%%%%%%%%%%%%%%%%%%%%%%%%%%%%%%%%%%%%%%%%%
%%%%%%%%%%%%%%%%%%%%%%%%%%%%%%%%%%%%%%%%%%%%%%%%%%%%%%%%%

\section{Improved treatment of gas accretion onto PBHs: The impact on the bounds}
\label{sec:improv_accretion}

The uncertainties associated with the physics of accretion play a major role in the estimation of the number of PBHs emitting above the threshold flux of successful detection. The constraints presented in \cite{Gaggero:2016dpq} were obtained based on the principles of weak black hole accretion, derived from our current understanding of black holes observed to accrete at very low rates, e.g., Sagittarius A*, the black hole at the Galactic center (see, e.g., \cite{Melia:2001,markoff:2010} for reviews on this black hole). Such low-level accretors lead to models for radiatively inefficient accretion flows (RIAFs), e.g. \cite{Narayan:1994}, and the non-linear scalings of radiative efficiency discussed in Section~\ref{subsec:intro_con} invoke these accretion scenarios. The accretion formalism of \cite{Gaggero:2016dpq} makes use of this radiative inefficiency scaling within the framework of Bondi-Hoyle-Littleton accretion. In this setup the accretion rate onto the isolated PBHs was parameterised by $\lambda=\dot{M}/\dot{M}_\mathrm{BHL}$, and this incorporated all our ignorance regarding the energetics of the gas as it travels from the Bondi radius to the inner regions of the accretion flow. Ref. \cite{Gaggero:2016dpq} then separately characterized the efficiency of converting the available gravitational energy into radiation through $\eta$, and factored in a two-phase gas regime, adjusting the sound speed according to whether or not the gas is likely ionized. Whilst this provides a reasonable framework within which to estimate the BH luminosity, a more physically-consistent approach would involve a more self-consistent estimate for how the accretion rate depends on the velocity of the black hole in the presence of radiative feedback. 

In the following sections we present an improved treatment of accretion onto our population of PBHs with radiative feedback.

% FIGURE 4 USED TO BE HERE ===========================================

\subsection{Radiation-hydrodynamic simulations}

In this section, we place our bound on more solid ground by implementing the results of a set of numerical simulations, presented in Ref. \cite{Park:2012cr}, aimed at characterizing the Bondi–Hoyle-Lyttleton problem---the accretion of baryonic matter onto an isolated, moving compact object---in the presence of significant radiative feedback. These results build on previous findings discussed in Ref.s \cite{Park:2010yh,Park:2011rf}, and are based on non-relativistic hydro-dynamic simulations that account for photo-heating and photo-ionization of the environment surrounding the compact object, due to the emission of UV and X-ray photons by the accretion disk and subsequent formation of an ionization front.

\subsection{Analytic formalism}

The fundamental distinction between the previous approach of simply parameterising the accretion rate in terms of some ratio of the Bondi-Hoyle-Littleton accretion rate, $\dot{M}_\mathrm{BHL}$, is that here we can instead prescribe the effective accretion rate within the ionized region behind the shock generated due to the motion of the BH. Thus here we may re-write the Bondi accretion formulae in terms of physical characteristics within this ionization region. 

Taking first the basic definition of $\langle{\lambda_\mathrm{B}}\rangle=\langle{\dot{M}}\rangle/\dot{M}_B$ as the mean accretion rate with respect to the Bondi rate (thus in contrast with the definition used by \cite{Gaggero:2016dpq} and shown in Equation~\ref{eq:old_accretion}, such that $\langle{\lambda_\mathrm{B}}\rangle=\lambda\dot{M}_\mathrm{BHL}/\dot{M}_\mathrm{B}$), Ref. \cite{Park:2012cr} finds
\begin{equation}
    \langle\lambda_\mathrm{B}\rangle = \frac{\rho_{\mathrm {in}}(GM_{\mathrm {BH}})^2}{(v^2_{\mathrm {in}}+c^2_{\mathrm {s, in}})^{3/2}}\frac{c^3_{\mathrm {s}}}{\rho(GM_{\mathrm {BH}})^2}=\frac{\rho_\mathrm{in}}{\rho}\Big(\frac{c_\mathrm{s}}{c_\mathrm{s,in}}\Big)^3\frac{1}{(1+v^2_\mathrm{in}/c^2_\mathrm{s,in})^{3/2}},
\end{equation}
{where the subscript `in' denotes the relevant quantities within the ionized region, while $\rho$ and $c_s$ are the ambient medium density and sound speed, respectively, as in Eq. \ref{eq:BHL_accretion}. }

%\begin{equation}
%    \langle\lambda_\mathrm{B}\rangle = \frac{\rho_{\mathrm {in}}(GM_{\mathrm {BH}})^2}{(v^2_{\mathrm {in}}+c^2_{\mathrm {s, in}})^{3/2}}\frac{c^3_{\mathrm {s, \infty}}}{\rho_{\mathrm 
%\infty}(GM_{\mathrm {BH}})^2}=\frac{\rho_\mathrm{in}}{\rho_\mathrm{\infty}}\Big(\frac{c_\mathrm{s,\infty}}{c_\mathrm{s,in}}\Big)^3\frac{1}{(1+v^2_%\mathrm{in}/c^2_\mathrm{s,in})^{3/2}},
%\end{equation}

%where the subscripts `in' and `$\infty$' denote the relevant quantities within the ionized region and in the ambient medium respectively. 

%Now we switch to dimensionless notation such that the Mach numbers, temperatures, and densities are given by $\mathcal M_\mathrm{in}\equiv v_\mathrm{in}/c_\mathrm{s,in}$ and $\mathcal M \equiv v_\mathrm{\infty}/c_\mathrm{s,\infty}$, $\Delta_T \equiv T_\mathrm{in}/T_\mathrm{\infty}$ and $\Delta_\rho \equiv \rho_\mathrm{in}/\rho_\mathrm{\infty}$, respectively. Since the gas within the ionized region is assumed to be fully ionized Hydrogen, the mean molecular weight of the gas decreases by a factor of 2, and thus $c^2_\mathrm{s,in}/c^2_\mathrm{s,\infty}=2\Delta_T$.  The accretion rate is thus 

{Now we switch to dimensionless notation such that the Mach numbers, temperatures, and densities are given by $\mathcal M_\mathrm{in}\equiv v_\mathrm{in}/c_\mathrm{s,in}$ and $\mathcal M \equiv v_\mathrm{BH}/c_\mathrm{s}$, $\Delta_T \equiv T_\mathrm{in}/T$ and $\Delta_\rho \equiv \rho_\mathrm{in}/\rho$, respectively.} 
Since the gas within the ionized region is assumed to be fully ionized Hydrogen, the mean molecular weight of the gas decreases by a factor of 2, and thus $c^2_\mathrm{s,in}/c^2_\mathrm{s}=2\Delta_T$.  The accretion rate is thus 
\begin{equation}
    \dot{M}=\langle\lambda_\mathrm{B}\rangle \dot{M}_B = \Delta_\rho (2\Delta_T)^{-3/2}(1+\mathcal M_\mathrm{in})^{-3/2}\frac{\pi \rho (GM_\mathrm{BH})^2}{c^3_\mathrm{s}}e^{3/2},
    \label{eq:mdot}
\end{equation}

where we have adopted the Bondi formula for an isothermal spherically accreting stationary BH, with the factor $e^{3/2}$ coming directly from the assumption of an isothermal gas. 

In order to calculate the accretion rate given by Equation~\ref{eq:mdot} we need explicit expressions for two terms: {\bf 1)} the Mach number within the ionized region, $\mathcal M_\mathrm{in}$, and {\bf 2)} the density ratio across the ionization shock front $\Delta_\rho$. 

The Mach number within the ionized region is related to the ambient Mach number by

%\begin{equation}
%    \mathcal M_\mathrm{in} = \frac{v_\mathrm{in}}{v_\mathrm{\infty}} \frac{v_\mathrm{\infty}}{c_\mathrm{s,\infty}} \frac{c_\mathrm{s,\infty}}{c_\mathrm{s,in}} = \frac{1}{\Delta_\rho}\mathcal M \frac{1}{(2\Delta_T)^{1/2}}. 
%    \label{eq:mdot_in}
%\end{equation}
\begin{equation}
    \mathcal M_\mathrm{in} = \frac{v_\mathrm{in}}{v_\mathrm{BH}} \frac{v_\mathrm{BH}}{c_\mathrm{s}} \frac{c_\mathrm{s}}{c_\mathrm{s,in}} = \frac{1}{\Delta_\rho} \mathcal M 
    \frac{1}{(2\Delta_T)^{1/2}}. 
    \label{eq:mdot_in}
\end{equation}

%By invoking both mass and momentum conservation, one can calculate the jump conditions for the density. 
%We refer the reader to \cite{Park:2012cr} for a detailed description of how the radiation-hydrodynamic simulation results are interpreted through the transitions from one type of ionization front (I-front) to another, and simply state that there exist two such fronts---rarefied, or $R$-type, and dense, or $D$-type. The key difference lies in the contrast in gas density across the I-front. An $R$-type I-front shows rough equilibrium between these two densities, $\rho_\mathrm{in} \sim \rho_\mathrm{\infty}$, whereas a $D$-type front exhibits a lower density downstream than the upstream density, $\rho_\mathrm{in} \le \rho_\mathrm{\infty}$. 

Regarding the latter term, we recall that---as detailed in \cite{Park:2012cr}---the radiation-hydrodynamic simulation results are interpreted through the transitions between two different types of ionization fronts as the BH mach number increases: rarefied, or $R$-type, and dense, or $D$-type. 
%{\bf The key difference lies in the contrast in gas density across the I-front: an $R$-type I-front shows rough equilibrium between these two densities, $\rho_\mathrm{in} \sim \rho_\mathrm{\infty}$, whereas a $D$-type front exhibits a lower density downstream than the upstream density, $\rho_\mathrm{in} \le \rho_\mathrm{\infty}$.}
{The key difference lies in the contrast in gas density across the I-front: an $R$-type I-front shows rough equilibrium between these two densities, $\rho_\mathrm{in} \sim \rho$, whereas a $D$-type front exhibits a lower density downstream than the upstream density, $\rho_\mathrm{in} \le \rho$.}
The full expressions as a function of the ambient Mach number and temperature ratio, which we will not derive here, are given by
\begin{equation}
    \Delta_\rho = \begin{cases} 
    \frac{1 + \mathcal{M}^2 + \sqrt{(1 + \mathcal{M}^2)^2 - 8\mathcal{M}^2\Delta_T}}{4\Delta_T}, & \mathrm{if} \,\, \mathcal{M} < \mathcal{M}_D, \\
    \frac{1 + \mathcal{M}^2}{4\Delta_T}, & \mathrm{if} \,\, \mathcal{M}_D \le \mathcal{M} \le \mathcal{M}_R, \\
    \frac{1 + \mathcal{M}^2 - \sqrt{(1 + \mathcal{M}^2)^2} - 8\mathcal{M}^2\Delta_T}{4\Delta_T}, & \mathrm{if} \,\, \mathcal{M} > \mathcal{M}_R. 
    \end{cases}
    \label{eq:density}
\end{equation}

Here the Mach number boundaries for $D$-type and $R$-type I-fronts are given by $\mathcal{M}_D=\sqrt{2\Delta_T}-\sqrt{2\Delta_T - 1}$ and $\mathcal{M}_R=\sqrt{2\Delta_T}+\sqrt{2\Delta_T - 1}$ respectively. Substituting Equations~\ref{eq:density} and~\ref{eq:mdot_in} into Equation~\ref{eq:mdot} allows us to fully characterize the PBH accretion rates in accordance with the numerical simulation results of \cite{Park:2012cr}. {Motivated by empirical values of the temperature in the CMZ (see \cite{2016MNRAS.457.2675H} and references therein, in particular \cite{Ao2013A&A}),} we adopt the same temperature assumptions as in \cite{Gaggero:2016dpq} {for the neutral and ionized regions respectively, $T=10^2$ K ($c_\mathrm{s} = 1$ km/s) and $T_\mathrm{in}=5\cdot10^3$ K ($c_\mathrm{s,in} = 10$ km/s)}, giving $\Delta_T=50$.

%We adopt the same temperature assumptions for the neutral and ionized regions respectively as in {\bf \cite{Gaggero:2016dpq} ($c_\mathrm{s}=1$~km/s and $c_\mathrm{s,in}=10$~km/s), giving $\Delta_T=50$. }

\begin{figure}[tb]
    \centering
    \includegraphics[width=\textwidth]{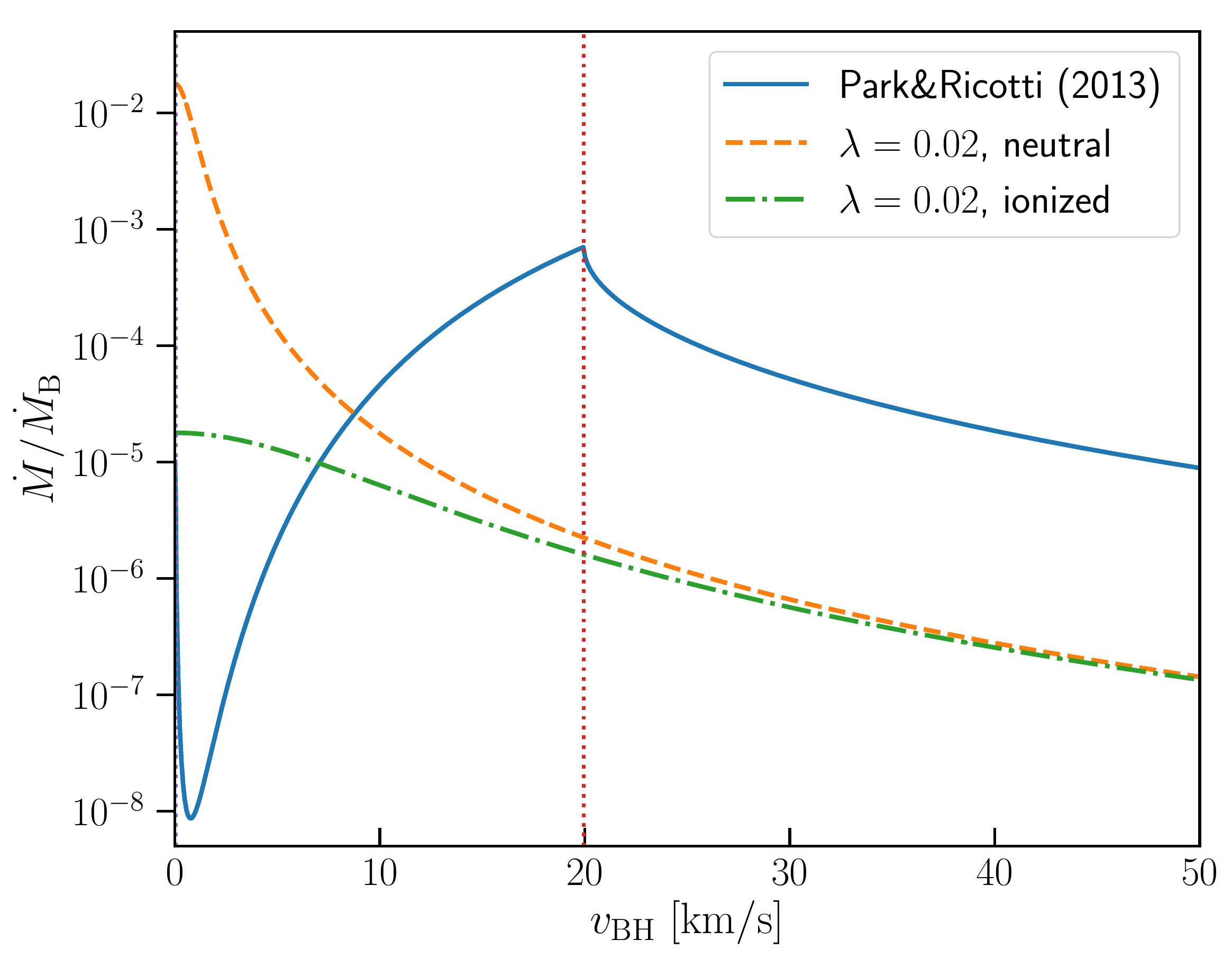}
    \caption{Accretion rates of interstellar gas onto a moving, isolated PBH as a function of its velocity. The results of \cite{Park:2012cr} (solid blue line) are  compared to the phenomenological prescription adopted in \cite{Gaggero:2016dpq}, whereby we show the two cases the authors considered: Bondi-Hoyle-Littleton accretion of neutral gas (orange dashed line) and gas which is considered fully ionized when the timescale for ionization is shorter than the timescale for the BH to traverse its Bondi sphere (green dot-dashed line). {The rates are expressed as fractions of the Bondi rate}, the mass of the PBH is fixed at $100 M_\odot$ and {the ambient gas density and temperature are set to $\rho = 10^4$ $m_p$ cm$^{-3}$ and $T = 10^2$ K respectively. The vertical dotted line identifies the Mach number where the accretion rate reaches the peak, i.e. where the ionization front starts to break down.}}
    \label{fig:park_ricotti_accretion}
\end{figure}

The accretion rate as a function of BH velocity is shown in Figure~\ref{fig:park_ricotti_accretion} (solid blue line), where we have compared the rates to the previous basic accretion assumptions of \cite{Gaggero:2016dpq} (dashed and dash-dotted lines). The plot clearly shows that the key feature of the new approach presented above and based on \cite{Park:2012cr} is the behaviour of the accretion rate as a function of the Mach number of the isolated moving black hole, captured by Eq. \ref{eq:mdot}. 

Let us comment further to convey a clearer physical intuition of this trend. While the classical Bondi–Hoyle–Lyttleton formula adopted in the previous work implies a simple, monotonic decrease of the accretion rate with increasing BH velocity, the simulations show a more complicated phenomenology. If the BH velocity is supersonic, but the Mach number is below a critical value $\mathcal{M}_R$ ($\mathcal{M}_R \simeq 4$ for $T \simeq 10^4$ K), a dense bow shock forms in the upstream region; behind the bow shock, a D-type (dense) ionization front develops, and a cometary-shaped HII region can be identified, characterized by low density and velocity. In this regime, the gas velocity in the reference frame of the moving BH decreases with increasing BH velocity, hence the accretion rate follows the opposite trend with respect to the Bondi-Hoyle-Lyttleton formula, i.e. it {\it increases with increasing BH velocity} (see Fig. \ref{fig:park_ricotti_accretion}).
Conversely, when the BH velocity is above $\mathcal{M}_R$, the ionization front becomes R-type (rarefied), and the accretion rate {\it decreases} with the BH velocity, effectively returning the accretion process to Bondi-Hoyle-Littleton-like accretion.

\subsection{Impact on the PBH constraints}

%Naturally, changing the accretion formalism influences the constraints as derived in \cite{Gaggero:2016dpq}. 
The accretion formalism described above has a noticeable impact on the constraints derived in \cite{Gaggero:2016dpq}. 
Upon further inspection of Fig. \ref{fig:park_ricotti_accretion}, it is possible to build an intuitive understanding for the constraints as a function of PBH mass and velocity. % understand this intuitively. 
With the new prescription (blue line), the accretion rates of PBHs with low velocities are significantly suppressed and should no longer contribute to the constraint; instead,  %inverted behavior of an increase in accretion rate with increasing BH speed, 
the accretion rate of PBHs with higher velocities is enhanced and therefore these objects are more likely to radiate above threshold. %This possibility is further strengthened by dropping the factor $\lambda=0.02$. 

Our simulations actually confirm this intuition and clearly show that %In fact, by reevaluating the constraints with the revised accretion formalism we find that 
the bound no longer originates from the PBHs at the low velocity tail, but from the PBHs with velocities around $2c_{s,\mathrm{in}} = 20$ km/s, i.e. near the peak of the accretion rate as plotted in Fig. \ref{fig:park_ricotti_accretion}.

\begin{figure}[b!]
\centering
\includegraphics[width=\textwidth]{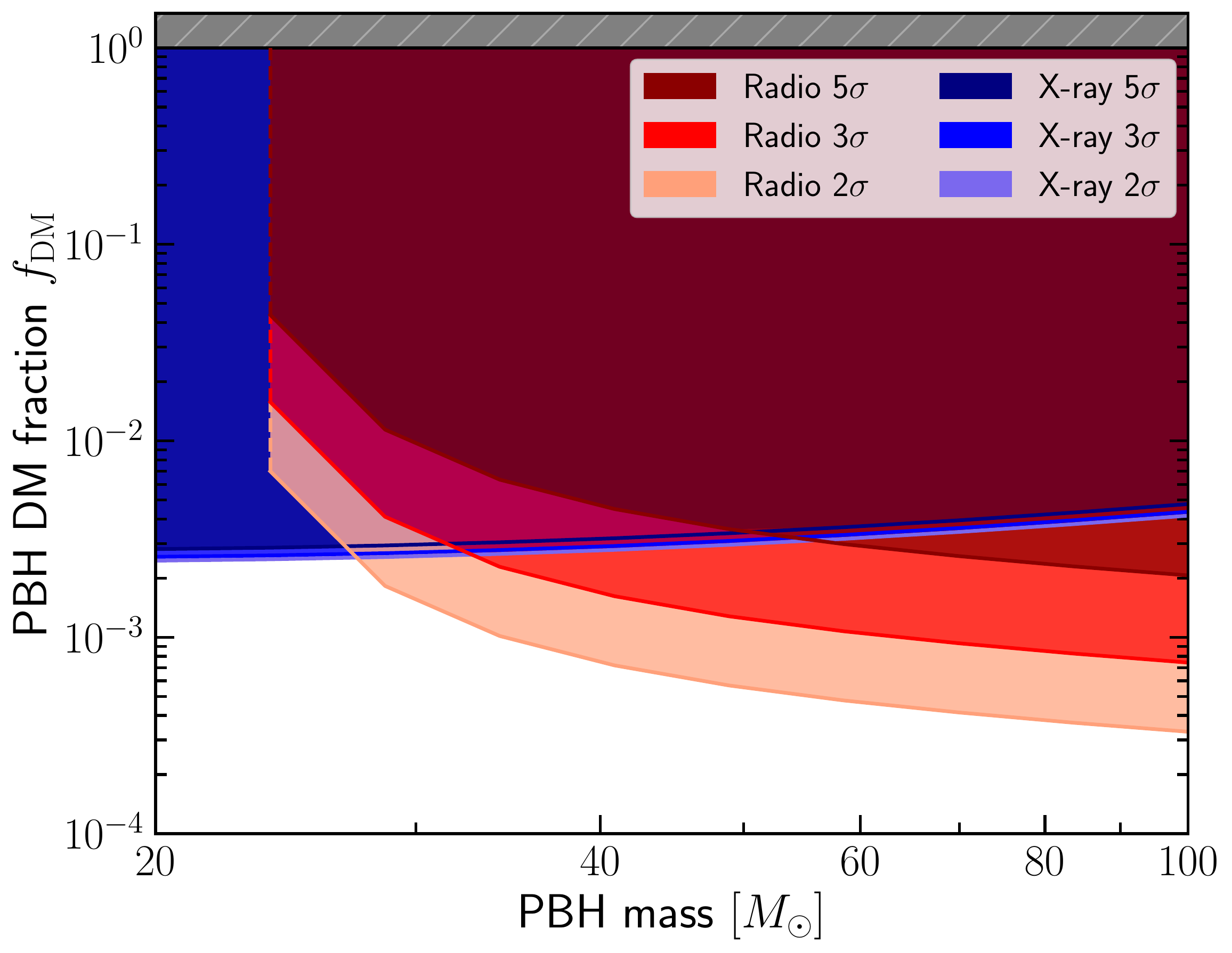}
\caption{Radio (red) and X-ray (blue) constraints on the PBH DM fraction as a function of PBH mass in the $20-100$ $M_\odot$ mass window. These constraints are obtained by implementing the revised accretion formalism from \cite{Park:2012cr}. 
The different lines correspond to different significance of exclusion: $2\sigma$ (bottom), $3\sigma$ (middle) and $5\sigma$ (top). The filled colors indicate the regions that are excluded with a certain significance, where lighter colors indicate lower significance. The gray region is unphysical, since the DM fraction exceeds 1.}
\label{fig:constraints_park2012}
\end{figure}

The resulting constraints are shown in Fig. \ref{fig:constraints_park2012}: They are two orders of magnitude stronger than those presented in \cite{Gaggero:2016dpq}, excluding PBH DM fractions above $\sim 10^{-3}$. This remarkable effect is mainly due to our result being sensitive to higher velocities---as pointed out above---which correspond to a larger fraction of the PBH population. %which can be attributed to a larger population of PBHs with the appropriate velocities: 20 km/s instead of less than a few km/s.
%Associated with these findings is also a different mass dependence, with the radio constraints at 20 $M_\odot$ vanishing completely. This feature can be explained by these PBHs not reaching sufficient accretion rates to be able to radiate above the radio detection threshold. %, i.e. by the peak at 20 km/s in Fig. \ref{fig:park_ricotti_accretion} not being high enough. 
%Apart from this threshold effect, both the radio and X-ray constraints are significantly stronger than those previously derived in \cite{Gaggero:2016dpq}. 

As far as the mass dependence is concerned, we notice that the radio and X-ray bound have different behaviors. 
This is due to two competing effects: 
{\bf 1)} On the one hand, the accretion rate rises with increases mass, increasing the probability of detection; {\bf 2)} on the other hand, the number density of PBHs decreases with increasing mass given the fixed reference dark matter mass density. 

For the radio bound, the former effect dominates since the constraining power is actually driven by a very small population of PBHs radiating close to the detection threshold, while the number of observed BH candidates in the radio is 0 as discussed in \cite{Gaggero:2016dpq}: therefore, the radio bound is not present for masses lower than 20 $M_\odot$, while for larger masses it ``kicks in'' and gets stronger with increasing mass. 

Concerning the X-ray bound---given the higher sensitivity of \textit{Chandra} and the resulting large population of PBHs radiating above threshold---the latter effect dominates: hence, the X-ray bound gets progressively weaker with increasing PBH mass. 

Finally, let us point out that the estimated radio luminosities of the PBH population, relative to their X-ray luminosity, also increases with BH mass in accordance with the FP scaling relation \cite{2012MNRAS.419..267P}, which also contributes to the relative strengthening of the radio bound at higher BH mass with respect to the X-ray bound.

\section{Discussion}

As a first discussion point, we want to emphasize the role of several critical aspects of accretion physics.
As pointed out in \cite{Park:2011rf} and \cite{Park:2012cr}, the flow of gas being accreted onto a moving, isolated compact object features a rich and complex phenomenology. 

An important feature is time dependence: According to simulations, at intermediate densities ($n_H \sim 10^3$ cm$^{-3}$, which is a close to the values measured in the CMZ clouds) the accretion rate shows an oscillatory behavior, and intermittent bursts can occur. These bursts are caused by an instability in the ionization shell in the upstream direction, and are seen to occur in radiation-hydrodynamic simulations in the Mach number range $2.5 \leq \mathcal{M} \leq \mathcal{M}_R$. The instability causes the shell to fragment and reform intermittently, with explosive accretion occurring while the shell is fragmented, and the reformation occurring in roughly the sound crossing time of the HII region. These bursts can increase the signal by a factor of 3 or more for a relatively short time (on the order of 1000 years, with a recurrence time of ~ 5000 years, both timescales depending on the gas temperature and density), and can significantly increase the probability of detecting a moving PBH that is crossing a dense cloud: we chose to neglect this effect, given the conservative approach we have followed, and the difficulty and the uncertainties involved in modeling this effect; however, it is useful to consider the fact that the bound can get even stronger if this behavior is taken into account. 

In addition, we recall that, while we are considering isolated black holes accreting gas as they traverse the CMZ, most of our knowledge on accreting BHs comes from those that are situated in X-ray binary systems: these are the closest such systems we can actually observe and study in detail. In this work, we have neglected the possibility that the more well-known outbursting behaviour of X-ray binaries could occur due to a critical mass forming in a disk-like structure during accretion (i.e. the disk instability model, see, e.g., \cite{Lasota:2001}). As mentioned by \cite{Park:2012cr}, the formation of an accretion disk-like structure akin to those found in the many accreting sources is not resolved in their simulations. 
As such we cannot be certain that such structures will not form around accreting isolated PBHs, and that this would not lead to the thermal instabilities expected in dense, thin accretion disks. The possibility of isolated BHs accreting from the ISM exhibiting such transient behaviour has been considered already \cite{Agol:2002,Matsumoto:2018}. If density perturbations within the CMZ exist on similar size scales to the Bondi sphere, the specific angular momentum of the gas may be sufficient for the gas to circularize prior to accreting \cite{Shapiro:1976}. Whilst \cite{Matsumoto:2018} suggest that gas densities in the accretion flows of isolated BHs can become sufficient to trigger Hydrogen ionization instabilities (and thus transient outbursts), as shown by \cite{Agol:2002}, the lack of detection of transient isolated systems (be they PBHs, astrophysical BHs, or Neutron stars) makes it likely that only a small fraction of isolated BHs actually exhibit outbursts of this nature. We have neglected this possibility, but note that folding in the predicted outburst rates of PBHs in the CMZ may have the effect of increasing the strength of the PBH DM bounds.  

Another key assumption behind the constraints derived here and in \cite{Gaggero:2016dpq} is the presence of jets launched from the accretion flows of the PBHs. As discussed by, e.g., \cite{Maccarone:2005}, radio searches for isolated BHs should be more successful than X-ray searches, since the jet kinetic power dominates the total accretion power with decreasing luminosity. This is a direct result of the observationally determined radio and X-ray flux correlations of X-ray binaries ($L_R\propto L_X^{0.7}$; \cite{Hannikainen:1998,Corbel:2000,Gallo:2003,Gallo:2014}), and is shown to be a mass-scaled relation \cite{Merloni:2003,heinz2003,Falcke:2004}, ultimately leading to the FP relation we use here to estimate the radio luminosities of our population of accreting PBHs \cite{2012MNRAS.419..267P}. As shown in Fig.~\ref{fig:constraints_park2012}, whilst the X-ray bounds on the PBH DM fraction are the most constraining at lower masses, at higher BH masses the radio bounds become stronger. This is due primarily to the vast difference in detected sources in the respective \textit{Chandra} and VLA Galactic center surveys, but is is also due in part to the mass-scaling and the dominance of jet kinetic power at the lowest accretion rates. With future radio telescopes such as SKA, we should expect the bounds to be dependent mostly on the radio surveys. Here, just as was done by \cite{Gaggero:2016dpq}, the predicted radio fluxes are based on the FP relation \cite{2012MNRAS.419..267P}. In order to apply the FP relation, we have to assume the presence of self-absorbed jets in our systems, which are emitting GHz radio
waves with an almost flat, optically thick spectrum. We remark that only the radio bound depends on this assumption, since the radio flux is estimated by means of a conversion from the X-ray band based on the FP relation. Ref. \cite{Park:2012cr} discuss the fact that the dynamical and radiative implications of the presence of jets is not considered in their simulations. 

The prevalence of jets in accreting BH systems across the mass scale at the lowest accretion rates (see, e.g., \cite{2012MNRAS.419..267P,Gallo:2014}) suggests that the structure of accretion flows (optically thin and geometrically thick) associated with low accretion rates are ideal for the launching of jets. This has been explored in the case of isolated BHs \cite{Armitage:1999, Maccarone:2005}, whereby the assumed conditions necessary for jet launching include a spinning BH and the presence of a magnetic field. We do not explore the likely distribution of PBH spins here, nor predict the strength of the magnetic field in the accretion flow, but we note that such conditions would be necessary. Thus the question of how this would impact the radiation-hydrodynamic simulations of \cite{Park:2012cr} is a caveat to our implementation in the simulations, but only for the derived radio bounds.

Another relevant source of uncertainty is the phase-space distribution of {both the visible and the dark matter in the inner part of the Galaxy.

Regarding the former, as mentioned e.g. in \cite{2014prpl.conf..125M,2016MNRAS.457.2675H} and references therein, an uncertainty of a factor of $\simeq 2$ or more about the total mass of the CMZ has to be taken into account.

As far as the latter is concerned,}
although there is evidence for the presence of dark matter inside the solar circle \cite{Iocco:2011jz}, the actual distribution at small Galactocentric radii is very uncertain. We assumed the NFW profile as a benchmark in this work, but we tested how the bound changes using different profiles. In particular, we remark that the scenario of PBHs composing all dark matter (i.e. $f_\mathrm{DM}=1$) is still ruled out at the $5\sigma$ confidence level even under the extreme assumption that the dark matter profile has a core as large as the size of the Galactic bulge.

%In particular, we remark that PBHs are still ruled out at the $5\sigma$ confidence level even under the extreme assumption that the dark matter profile has a core as large as the size of the Galactic bulge. 
%the bound is still present under the very conservative assumption of a flat dark matter profile, despite weakening significantly by 1-2 orders of magnitude. %[This indicates that... the DM profile cannot kill the constraints?] {\tt put here the test with the extremely conservative profile... done?}

As a further validation of our results, we considered an improved estimate of the velocity distribution %{As far as the velocity distribution is concerned, as a further validation of our results we considered an improved estimate}
based on the Eddington formalism \cite{BinneyTremaine}. Under the assumption of spherical symmetry and isotropy, the Eddington inversion formula allows us to compute the velocity distribution at different Galacticentric radii $f(r,v)$ given a density profile $\rho(r)$. This formalism has been successfully applied in the context of the determination of the DM density in the Solar neighborhood \cite{Catena:2011kv}. In our case, since the constraining power mainly comes from the very inner regions of the Galaxy, the assumptions behind the formula are put on even safer grounds. 
Moreover, while in the previous paper the constraining power in the momentum space mainly came from the very-low-velocity tail of the distribution, now---given the increase of the accretion rate with the PBH velocity---PBHs above the detection threshold cover a wide range of velocities. Although the impact of the Eddington formalism, which departs from the Maxwell-Boltzmann approximation especially at large speed, may thus be in principle significant, we have verified that it has little impact on the X-ray and radio bound. %{\tt Quantify...}

Regarding the velocity distribution, we stress that the velocity we are interested in is the relative one between the black hole and the gas, which can be affected by the turbulent motion of the gas itself. This point is stressed by the authors of \cite{Hektor:2018rul} who revisited the bound proposed in our previous work by adding the turbulent speed to the denominator of {the accretion rate (i.e., to the denominator of Eq. \ref{eq:BHL_accretion}, substituted into Eq. \ref{eq:old_accretion})} as follows: $\dot{M}=4\pi\lambda (G M_{BH})^2 \rho \left(v_{BH}^2+v_{turb}^2\right)^{-3/2}$. However, they have overestimated the impact of turbulent motion by adopting as a reference value for $v_{turb}$---taken from \cite{2016MNRAS.457.2675H} and based on spectral line observations of molecules such as HNCO, N2H+, and HNC---an upper limit, not an estimate, of the speed of turbulent motion \cite{2016MNRAS.457.2675H}.
Even assuming the maximum value for $v_{turb}$, our results do not change significantly, since the bulk of the high-luminosity radio and X-ray sources, as we stressed several times above, do not arise from PBHs in the very-low velocity tail.

As a further caveat, let us mention that there has been a debate in the literature about the clustering level of PBHs \cite{Chisholm:2005vm,Desjacques:2018wuu,Ballesteros:2018swv,Bringmann:2018mxj} and its impact on the bounds. Whilst a quantitative estimate of this effect is beyond the scope of the present work, we remark that a significant clustering could in principle go in the direction of strenghtening the bound, since a cluster of PBHs crossing a dense gas cloud would be more easily detected, and leave a significant and observable imprint on the environment due to the localized, high flux of ionizing photons from the accreted gas.
%In conclusion, let us remark that the radio and X-ray channels can be considered as an actual window of detection with enormous potential, in light of the planned SKA observatory which is expected to provide a dramatic increase in sensitivity. This window of detection of course applies to the guaranteed population of astrophysical black holes as well.
%The future challenge, from an even broader perspective, consists then in devising effective strategies to interpret the current and forthcoming multi-messenger data in order to detect, in different contexts, a population of potential PBHs, and distinguish it from an expected ordinary population of astrophysics black holes.

As a final point, we stress that PBHs searches in the radio are particularly promising, in view of the upcoming radio facilities, including the SKA observatory \cite{Bull:2018lat} and ngVLA \cite{Murphy:2018vxa}, which are expected to provide a very significant increase in sensitivity. 
One of the main challenges will be then to devise effective multi-wavelength strategies to disentangle accreting PBHs from other radio sources, and in particular from the guaranteed population of astrophysical black holes, since the current bound is approaching the estimated density associated with the population of stellar-mass astrophysical black holes in our Galaxy that form through conventional stellar evolution.
Such a population could amount to $\sim 10^8$, see e.g. \cite{1983PhT....36j..89S,Fender:2013ei}: this estimate stems from a modeling the star formation history of the Milky Way and our knowledge about the local mass density of stellar remnants.
Very little is currently known about this population, with a few exceptions (namely, the X-ray binaries---i.e. the black holes that are accreting mass from a companion star---and possibly a black hole candidate part of a wide binary system recently announced in \cite{Thompson:2018ycv}).

\section{Conclusions}

In this paper we have discussed astronomical searches for primordial black holes based on the analysis of radio and X-ray data from the Galactic center region. We have updated the contraints on the abundance of PBHs presented in \cite{Gaggero:2016dpq} by taking into account the impact of an extended mass function and by introducing a more detailed treatment of the physics of gas accretion onto moving PBHs. 
We have shown that, in general, extended mass functions lead to constraints that are more stringent by a factor of a few with respect to the case of a monochromatic mass function. We have validated our results via a semi-analytical remapping procedure.

As for the physics of gas accretion, we have implemented, for the first time in this context, the results of a set of state-of-the-art numerical simulations that capture the rich phenomenology of gas accretion onto a moving compact object in the presence of radiative feedback. This new approach allowed us to obtain a strong bound on the PBH abundance, which is at the level of $\sim 10^{-3}$ of the dark matter density (and close to the expected level for the astrophysical black hole density). The bound comes from both X-ray and radio data, and 
%is based on solid results regarding the phenomenology of accretion; 
with respect to the previous upper limit reported in \cite{Gaggero:2016dpq}, it appears less sensitive to the details of the low-velocity tail of the velocity distribution.

In conclusion, astronomical data strongly disfavor primordial black holes as the main constituent of the dark matter in the mass range under consideration, and can be considered a promising path of detection for both a sub-dominant population of those objects (with a number density within the allowed range), and the expected population of astrophysical black holes, either isolated or as part of wide binary systems, the detection of which is a current challenge of modern astrophysics.

\section*{Acknowledgements}

We thank G. Ballesteros, J. Garcia-Bellido, and T. Maccarone for important comments and suggestions.
D.G. has received financial support during the last stage of this project through the Postdoctoral Junior Leader Fellowship Programme from la Caixa Banking Foundation. R.M.T.C. acknowledges
support from NASA Grant No. 80NSSC177K0515.

\newpage
\begin{appendix}

\section{Derivation of $g(M,\{p_j\})$}
\label{app:gfunction}

As mentioned in Sec. \ref{subsec:remap}, the observable of interest is the number of PBHs radiating above the detection threshold. With each PBH contributing independently to this number, the observable is proportional to Eq. \ref{eq:observable}. Therefore, the number of sources above the detection threshold $N_S$, expressed as an integral over the flux distribution of the PBH population $dN/d\phi$, can be equated to an integral over the mass distribution,
\begin{equation}
N_S = \int_{\phi_\mathrm{thres}}^\infty \frac{dN}{d\phi} \, d\phi = \int_0^\infty g(M,\{p_j\}) \frac{dN}{dM} \, dM. \label{eq:obsvsg}
\end{equation}
Here the proportionality between the observable and Eq. \ref{eq:observable} has been fixed by the usage of the total mass distribution $dN/dM \equiv N_\mathrm{tot}\, d\Phi/dM$, with the total number of PBHs $N_\mathrm{tot}$ having an implicit dependence on the PBH dark matter fraction $f_\mathrm{PBH}$. Now, an intuitive description of $g$ becomes visible; it is the fraction of PBHs with mass $M$ that contribute to the number of sources with emission above the detection threshold.

In principle, the flux detected at Earth originating from an accreting PBH depends on many different parameters. Within the simulations, this is limited to three randomly chosen numbers: the PBH mass, the PBH velocity and its position. To obtain the flux distribution, one needs to convert the mass, velocity and position distributions following the calculations performed in the simulations. Given the complexity introduced by the random locations, we proceed only with a random velocity and mass. Therefore, we are making the assumptions of a constant velocity distribution, distance and gas density.

{For simplicity, we also make the assumption of a bijective mapping of the BH velocity and mass to its radiative flux. Clearly, this assumption is only applicable to the accretion rate considered in Sec. \ref{sec:impact_mass}, and not to that considered in Sec. \ref{sec:improv_accretion} where multiple velocities can correspond to the same accretion rate (see Fig. \ref{fig:park_ricotti_accretion}). One could in principle extend the following formalism also for Sec. \ref{sec:improv_accretion} by introducing thoroughly chosen breaks in the velocity integration limits, but this is beyond the scope of this paper. Thus the remainder of the Appendix only applies to Sec. \ref{sec:impact_mass}.

The aforementioned assumption allows us to perform} a change of variables on the flux distribution in Eq. \ref{eq:obsvsg}. {By consequently} comparing it with the right-hand side, an implicit expression for $g$ can be obtained,
\begin{equation}
\nonumber
g(M,\{p_j\}) = \int_{\phi_\mathrm{thres}}^\infty p(h(\phi,M)) \left| \frac{\partial h(\phi,M)}{\partial \phi} \right| \, d\phi .
\end{equation}
Here $p$ is the normalized velocity distribution, and the function $h(\phi,M)$ maps a given flux $\phi$ and mass $M$ to the velocity required for a PBH with mass $M$ to obtain this flux. Given that {under our assumption} the flux is a monotonically decreasing function of the velocity, $g$ is nothing more than the CDF of the velocity distribution, properly taking into account the detection threshold. Since we are considering a Maxwell-Boltzmann velocity distribution, a more explicit version of $g$ is given by its CDF,
\begin{equation}
\nonumber
g(M,\{p_j\}) =
\begin{cases}
0,  & \,\, \mathrm{if} \: M \leq M_\mathrm{min},\\
\mathrm{erf}\left(\frac{h_\mathrm{thres}}{\sqrt{2}a}\right)-\sqrt{\frac{2}{\pi}}\frac{h_\mathrm{thres}}{a}\,\mathrm{exp}\left(-\frac{-h_\mathrm{thres}^2}{2a^2}\right), & \,\, \mathrm{if} \: M > M_\mathrm{min},
\end{cases}
\end{equation}
with the mass dependence hidden in $h_\mathrm{thres}\equiv h(\phi_\mathrm{thres},M)$. The minimum mass $M_\mathrm{min}$ is introduced due to the fact that in the case of a steady black hole the accretion is limited by the sound speed of the accreting gas, and thus light black holes may never reach fluxes above the detection threshold.

The last step is to find an explicit expression for the function $h$, containing the calculation of the flux from the mass and velocity as in the simulations. This depends on the flux of interest and are given by
\begin{equation}
\nonumber
h(\phi_R,M) = \left(\phi_R^{-1.45/3} X M^{3.88/3} - c_s^2\right)^{1/2} \quad \mathrm{and} \quad h(\phi_X,M) = \left(\phi_X^{-1/3} Y M - c_s^2 \right)^{1/2} 
\end{equation}
for the 1.4 GHz radio and $0.5-8$ keV X-ray flux respectively. Here $c_s$ is the sound speed of the surrounding medium, and $X=0.0448 \, (\mathrm{mJy})^{1.45/3} \, (M_\odot)^{-3.88/3} \, (\mathrm{km/s})^2$ and $Y=0.0277 \, (\mathrm{ph} \, \mathrm{cm}^{-2} \mathrm{s}^{-1})^{1/3} \, (M_\odot)^{-1} \, (\mathrm{km/s})^2$ are constants.

\end{appendix}

\bibliographystyle{JHEP}
\bibliography{PBHbiblio}

\end{document}